\documentclass{ws-ijmpd}
\usepackage{epsfig}
\usepackage{epstopdf}
\usepackage{color}

\begin{document}

\title{A study on the effect of anisotropy under Finch-Skea geometry} 

\author{Shyam Das}
\address{Department of Physics, Malda College, Malda 732101, West Bengal, India \\
dasshyam321@gmail.com}

\author{Koushik Chakraborty}
\address{Department of Physics, Government College of Education, Burdwan 713102, West Bengal, India \\koushik@associates.iucaa.in}

\author{Lipi Baskey}
\address{Department of Mathematics, Government General Degree College at Kushmandi, Dakshin Dinajpur 733121, West Bengal, India\\
lipibaskey@gmail.com}

\author{Saibal Ray}
\address{Department of Physics, Government College of Engineering and Ceramic Technology, Kolkata 700010, West Bengal, India\\
saibal@associates.iucaa.in}

\maketitle	

\date{Received: date / Accepted: date}

\abstract{The popularity of the Finch-Skea ansatz~\cite{Finch} to describe relativistic stellar model have encouraged us to study the analytic solutions of the Einstein field equation. We have presented a class of exact solutions to the field equations after considering the corresponding two cases: (i) positive value of anisotropic parameter, and (ii) absence of any anisotropy. Smooth matching of the interior solutions with the Schwarzschild exterior solution helped us to determine constants. The physical features of the solutions thus obtained have been studied both graphically and numerically with the specific pulsar $4U~1608-52$ (Mass = $1.57^{+0.30}_{-0.29}~M_\odot$ and radius = $9.8\pm 1.8~km$~\cite{roupas}). The stability conditions for the model have also been discussed, however the model is found to be stable for zero anisotropy.}


\section{\label{sec1} Introduction}
Exact analytic solutions of General Relativistic field equations is an active field of study for over a century, beginning its journey from the first ever solution proposed by Schwarzschild~\cite{Schwarzschild}. On this aspect some thorough reviews were published from time to time~\cite{kramer}. In particular, static spherically symmetric perfect fluid solutions were reviewed by Delgaty and Lake~\cite{Delgaty}. On close examination, they reported $16$ static spherically symmetric perfect fluid solutions to be physically viable out of total $127$ considered in the review. Correcting the solution given by Duorah and Ray~\cite{Duorah}, Finch and Skea~\cite{Finch} obtained a new exact solution to the Einstein field equations. One astonishing feature of the solution is that their model described compact star in isotropic pressure only. Therefore, here motivation for the present work is to consider a model of anisotropic star that reduces to the Finch-Skea solution for zero anisotropy.

The Finch-Skea metric~\cite{Finch} is well behaved and it satisfies all the criteria to be a Delgaty and Lake solution~\cite{Delgaty} for a static spherically symmetric perfect fluid model and also has been shown to be consistent with the Walecka theory~\cite{Walecka74} for cold condensed star. Additionally, this metric is found to be consistent to study neutron star, especially to investigate central densities of neutron star in relativistic mean-field theory~\cite{Walecka75}. Essentially this spacetime satisfies the characteristic of a perfect fluid matter which obeys barotropic equation of state (EOS). The underlying approach to solving Einstein's equations for spherical symmetry involved {\it ad hoc} assumptions for one of the gravitational potentials as the system of field equations is under-determined and possesses one degree of freedom. It is interesting to note that the Finch-Skea metric involves theorizing a form for the radial potential which allows for the complete integration of the field equations whereupon all the remaining geometric and dynamical quantities may be determined~\cite{CH15}. Kalam et al.~\cite{Kalam13a,Kalam13b} have invested their time to model strange quark stars using the Finch-Skea metric considering the MIT bag model~\cite{Kalam13a} as well as two fluid model~\cite{Kalam13b} and have proposed quintessence stars combining anisotropic pressure corresponding to normal matter~\cite{Kalam13c}. 

On the other hand, Maharaj et al.~\cite{Maharaj16} have shown that the Finch-Skea geometry can be generalized to include charge and anisotropy. Considering a particular solution to the charged anisotropy of this model Kileba et al.~\cite{Matondo17} have predicted the masses of the stellar objects for three different scenarios, viz, (i) charged anisotropic, (ii) charged isotropic, and (iii) uncharged isotropic distributions which were found to be compatible with several known compact objects. Tikekar and Jotania~\cite{TJ07} applied the Finch-Skea metric~\cite{Finch}, by assuming the $3$-space of the interior spacetime of a strange star is that of a three-paraboloid immersed in a 4-dimensional Euclidean space, to acquire a two-parameter family of physically viable relativistic models of neutron stars and showed that it admitted possibilities of describing strange stars as well as other highly compact configurations of matter in equilibrium. The Finch-Skea ansatz~\cite{Finch} was also used Sharma and Ratanpal~\cite{SR13} to generate a class of solutions describing the interior of a static spherically symmetric anisotropic star. Later, Pandya et al.~\cite{PTS} have generalized the model of Sharma and Ratanpal~\cite{SR13} by incorporating a dimensionless parameter $n(>0)$ in the Finch-Skea ansatz~\cite{Finch} by assuming the system to be anisotropic. Charged Finch-Skea stars were described in terms of Bessel functions and modified Bessel functions, by Hansraj and Maharaj~\cite{HM06} and Hansraj et al.~\cite{HM16}, where both the models are found to obey a barotropic EOS.

Bhar et al.~\cite{Bhar14} have produced anisotropic stars in ($2 + 1$) dimensions and a quark EOS by using the Finch-Skea metric~\cite{Finch}.  A class of interior solutions corresponding to the BTZ~\cite{BTZ} exterior solution has been investigated by Banerjee et al.~\cite{Banerjee13} under the Finch-Skea metric which is relevant for the description of realistic stars in ($3 + 1$) dimensions as a complementary approach to the study by Garc{\'i}a et al.~\cite{Garcia03}.

For higher dimensions, several researchers~\cite{Hansraj17,Dadhich17,Molina17,Patel97,CH15} have studied the Finch-Skea metric as well as its generalizations. Another fascinating study by Hansraj et al.~\cite{Hans15} shows that the Finch-Skea spacetime also arises in the $5$-dimensional Einstein-Gauss-Bonnet modified theory of gravity, suggesting that the Finch-Skea geometry may play an important role in more general Lovelock polynomials with a Lagrangian containing higher order terms~\cite{Maharaj16}.  

Now, in the present model the radial pressure is taken different from the transverse components of the pressure and thus anisotropy implies unequal principal stresses. Equality of the transverse components of pressure ensures the spherical symmetry of the model~\cite{Gleiser2002}. Various reasons are being pointed out by the researchers as the origin of anisotropy inside the compact star. Anisotropy can develop in the core of the compact stars due to exotic phase transition at extreme density~\cite{Sokolov}. Jones~\cite{Jones} predicted the presence of type II superconductor inside the compact stars leading to the anisotropy of stress tensor. Pion condensation~\cite{Sawyer}, type $3A$ superfluid~\cite{Kippen} are also identified as the possible origins of anisotropy. Ruderman~\cite{Ruderman} indicated that local anisotropy may develope in compact stars due to the solid core. Strong magnetic field may also lead to the development of anisotropy inside the compact star~\cite{Weber}. Liebling and Palenzuela~\cite{Liebling} have shown that a scalar field in a Boson star may give rise to anisotropy. However, one can look into for review of the local anisotropy by Herrera and Santos~\cite{Herrera97}.

It is worthy to note that anisotropy in pressure plays a significant role in the structure and properties of compact star. Karmarkar et al.~\cite{Karmarkar} indicated that numerical value of the compactness parameter $\frac{2M}{R}$, $M$ and $R$ being the mass and radius of the star, may approach unity for anisotropic stars. The upper limit of the surface redshift for anisotropic stars becomes $3.842$ and $5.211$ when the transverse components of the pressure satisfy the strong and the dominant energy condition respectively~\cite{Ivanov2002}. Mak and Harko~\cite{Mak1,Mak2} have showed that anisotropy must be maximum at the surface of the compact star and it should be zero at the center of the fluid sphere. There are good number of models on anisotropic compact star under General Relativity in literature~\cite{Gleiser2003,Rahaman1,Varela,Sharma7,Maharaj}. In the present paper we have investigated the nature of anisotropy to model a stable compact star. 
 
The structure of the paper is as follows: in Sec. \ref{Sec2} the basic equations required for the description of the model of the compact star are presented. The relevant solutions to the Einstein field equations for the model are presented in Sec.~\ref{Sec3} by discussing two cases:  (i) anisotropic case and (ii) isotropic case. The smooth matching for the interior and the exterior solution at the boundary is discussed in Sec.~\ref{Sec4} and hence the general form for the integration constants are found. In Sec. \ref{Sec5} we elaborate the physical analysis for a viable model of compact star. We have discussed the stability analysis for the prescribed model in Sec.~\ref{Sec6}. The mass-radius relationship, and surface redshift for the model are described in Sec. \ref{Sec7}. Additionally, the variation of the central density with the mass and the radius are discussed in Sec. \ref{Sec7}. The last section is dedicated to concluding remarks.

\section{\label{Sec2} Einstein's field equation and the interior solution for the model}
We start by considering the model which represents a static spherically symmetric fluid configuration. The line element describing the interior space-time of a spherically symmetric star in Schwarzschild coordinates\\
$x^0 = t$,  $x^1=r$,  $x^2 = \theta$,  $x^3 = \phi$ can be written as
\begin{equation}
ds_{-}^2 = -A_{0}^2(r)dt^2 + B_{0}^2(r)dr^2 + r^2(d\theta^2 + \sin^2\theta d\phi^2),\label{eq1}
\end{equation}
where $A_{0}(r)$ and  $B_{0}(r)$ are the gravitational potential yet to be established.

To study stellar structure and stellar evolution the basic supposition made by the researchers is to consider the interior of a star as a perfect fluid~\cite{DDC,Kippen}. The pressure in the interior of a star is considered to be isotropic to model this perfect fluid \cite{Gleiser2002}. Several studies in recent times have shown that, at very high density, alteration in isotropic pressure plays a vital role in studying the features of a stellar interior~\cite{Ruderman,Canuto}. Thus the energy momentum tensor is anisotropic, isotropy being the extra assumptions on the behaviors of the fields or of the fluid modeling the stellar interior~\cite{Gleiser2002}.

The matter distribution of the stellar interior is thus described by an energy-momentum tensor of the form
\begin{equation}
T_{\alpha\beta} = (\rho + p_t)u_{\alpha} {u_\beta} + p_{t} g_{\alpha \beta} + (p_r - p_t)\chi_{\alpha} \chi_{\beta},\label{eq2}
\end{equation}
where $\rho$ represents the energy-density, $p_r$ and $p_t$, respectively denote fluid pressures along the radial and transverse directions, $u^\alpha$ is the $4$-velocity of the fluid and $\chi^\alpha$ is a unit space-like $4$-vector along the radial direction so that $u^\alpha u_\alpha = 1$, $\chi_\alpha \chi_\beta =-1$ and $u^\alpha\chi_\beta=0 $.

The Einstein field equations governing the evolution of the system is then obtained as (we set  $G = c = 1$)
\begin{eqnarray}
8\pi\rho &=& \left[\frac{1}{r^2}-\frac{1}{r^2 B_0^2}+\frac{2B_0'}{r B_0^3}\right],\label{feq3}\\
8\pi p_r &=& \left[-\frac{1}{r^2}+\frac{1}{B_0^2 r^2}+\frac{2A_0'}{r A_0B_0^2}\right],\label{feq4}\\
8\pi p_t &=& \left[\frac{A_0''}{A_0B_0^2} + \frac{A_0'}{rA_0B_0^2} - \frac{B_0'}{r B_0^3} - \frac{A_0'B_0'}{A_0 B_0^3}\right],
\label{feq5}
\end{eqnarray}
where in above set of Eqs.~(\ref{feq3})-(\ref{feq5}), a `prime' denotes differentiation with respect to $r$.

Making use of Eqs.~(\ref{feq4}) and (\ref{feq5}), we define the anisotropic parameter of the stellar system as
\[\Delta(r) = 8\pi (p_t-p_r) = \]
\begin{equation}\label{aeq6}
\left[\frac{A_0''}{A_0B_0^2} - \frac{A_0'}{r A_0B_0^2} - \frac{B_0'}{r B_0^3} +\frac{A_0'B_0'}{A_0B_0^3} - \frac{1}{r^2B_0^2} + \frac{1}{r^2}\right].
\end{equation}

Moreover, the mass contained within a radius $r$ of the sphere is defined as
\begin{equation}
m(r)= \frac{1}{2}\int_0^r\omega^2 \rho(\omega)d\omega.\label{eq7}
\end{equation}

At this stage, we have a system of equations consisting of three equations Eq.~(\ref{feq3})- Eq.~(\ref{feq5}) with five unknowns  $\rho$, $p_r$, $p_t$, $A_0(r)$ and $B_0(r)$. Thus to find the exact solutions of the field equations and hence to model a stellar interior, we need to specify two of them. To model a physically reasonable stellar configuration, we propose that the metric potential $g_{rr}$ is of the form as considered by Finch-Skea~\cite{Finch} and it is given by
\begin{equation}
B_0^2 (r) = \left(1 +\frac{r^2}{R^2}\right),\label{eq12b}
\end{equation}
where $R$ is the curvature parameter describing the geometry of the configuration having a dimension of length. This choice of metric potential assures that the function $B^2_0(r)$ is finite, continuous and well defined within stellar interior range. Also $B^2_0(r)=1$ for $r=0$ ensures that it is finite at the center. Again, the metric is regular at the center since $(B^2_0(r))'_{r=0}=0$. 

With this choice of $B_0(r)$, Eq.~(\ref{aeq6}) then reduces to
\begin{equation}\label{eq11}
\Delta(r) = \frac{r^3 A_0 + R^2 \left[r \left(r^2 + R^2 \right) A_0'' - \left(R^2 + 2r^2 \right) A_0' \right]}{r \left(r^2 + R^2 \right) A_0}.
\end{equation}

On rearranging Eq.~(\ref{eq11}), we get
\begin{equation}\label{eq12}
\frac{R^2 A_0''}{A_0} - \frac{R^2 \left(R^2 + 2 r^2 \right)}{r \left( r^2 + R^2 \right)} \frac{A_0'}{A_0} + \frac{r^2}{r^2 + R^2} = \Delta (r).
\end{equation}

Now the above Eq.~(\ref{eq12}) can be solved for $A_0(r)$ if the anisotropic parameter, $\Delta(r)$ is specified in particular form. One can easily obtain solutions for the following two cases from Eq.~(\ref{eq12}): (i) $\Delta(r) = 0$ and (ii) $\Delta(r) \neq 0$. We have discussed both the cases in the next Section.

\section{\label{Sec3} Exact solutions to field equations }
To obtain an exact solution for Eq.~(\ref{eq12}), we need to consider the anisotropy in some specific form. We therefore consider the the anisotropic factor as
\begin{equation}
\Delta = \frac{\alpha r^2 (R^2 - r^2)}{(R^2 + r^2)^3}, \label{delalpha}
\end{equation}
where $\alpha$ is the parameter determining the measure of the anisotropy. Now this choice of anisotropy is feasible for the consideration of the anisotropic factor as $\Delta$ is regular for the radial coordinate $r$ and also $\Delta (r=0)=0$ is satisfied at the center. Utilizing this choice of anisotropy in Eq.~(\ref{eq12}), we get the master equation in the form\\
\[ \frac{\big(r R^2 (r^2 + R^2)^3 \big)A''_0 - \left(R^2 (2 r^2 + R^2)(r^2 + R^2)^2\right)A'_0}{r(r^2 + R^2)^3 A_0} \]
\begin{equation}
+ \frac{r^3 \left((r^2+R^2)^2 + \alpha (r^2 - R^2)\right)A_0}{r(r^2 + R^2)^3 A_0}=0.\label{eqme}
\end{equation}

Now we would like to study the exact solutions of the field equations for different values of $\alpha$. In the present work, we have investigated the values of $\alpha$ for $-1,~0$ and $1$. It is worth mentioning that similar investigation have been conducted by Sharma and Das~\cite{SD13} using the Finch-Skea metric. Their model represents an initially static star which is either anisotropic or isotropic in nature and which eventually describes a gravitationally collapsing system. However, in the present work we are attempting to depict a spherically symmetric stable configuration, for both anisotropic and isotropic nature of pressure.

Now using Frobenius Method~\cite{Teschl2012} we can solve Eq.~(\ref{eqme}) at $r~=~0$. Considering the solution in the series form, it can be written as
\begin{equation}
A_0 = \sum_{n=1}^{\infty} c_n r^{s+n},~ c_0 \neq 0.\label{eqc0}
\end{equation}

Computing the values of $A'_0$ and $A''_0$ and substituting all the values on Eq.~(\ref{eqme}), we obtain the following differential equation as
\[ r R^2 (r^6+3 r^4 R^2 + 3 r^2 R^2 + R^6)\]
 \[\sum_{n=1}^{\infty} (s+n)(s+n-1)c_n r^{s+n-2} \]
 \[ - R^2 (2 r^6 + 5 r^4 R^2 + 4 r^2 R^4 + R^6) \sum_{n=1}^{\infty} (s+n) c_n r^{s +n-1}\]
\begin{equation}
+ r^3 (r^4 + 2 r^2 R^2 + R^4 + \alpha r^2 -  \alpha R^2) \sum_{n=1}^{\infty}c_n r^{s+n} = 0. \label{eqde}
\end{equation}

Equating to zero the coefficient of the smallest power of $r$ and hence solving, we get the roots of  the indicial equation as $0$ and $2$. Further solving for each coefficient of $r$ we get the solution of Eq.~(\ref{eqme}) as
\begin{equation}
A_0 = M u(r) + N v(r), \label{eqsol1}
\end{equation}
where $M$ and $N$ are two arbitrary constants and 
\begin{eqnarray}
u(r) &=& 1 + {7 \over 2R^2} r^2 + \frac{6 R^2 + \alpha}{8 R^6} r^4 +... \\ \nonumber
v(r) &=& r^2 \left[ 1+ \frac{1}{4 R^2} r^2 + \frac{\alpha - 2 R^2}{24 R^6} +...  \right]. \label{eqsol2}
\end{eqnarray}

However due to complexity of the solution we investigate the exact solution of the model by considering specific values of $\alpha$ which is described in the following subsections.

\subsection{Exact solution in the presence of anisotropy} 

\subsubsection{Case I: $\alpha=-1$}
To obtain an exact solution to Eq.~(\ref{eq12}), we assume the anisotropic parameter to be in the form
\begin{equation}
\Delta(r)= \frac{r^2 \left(r^2 - R^2 \right)}{\left(r^2 + R^2 \right)^3}.\label{eq13}
\end{equation}

\begin{figure}[!htbp]
\begin{center}
\begin{tabular}{lr}
\includegraphics[width=8cm]{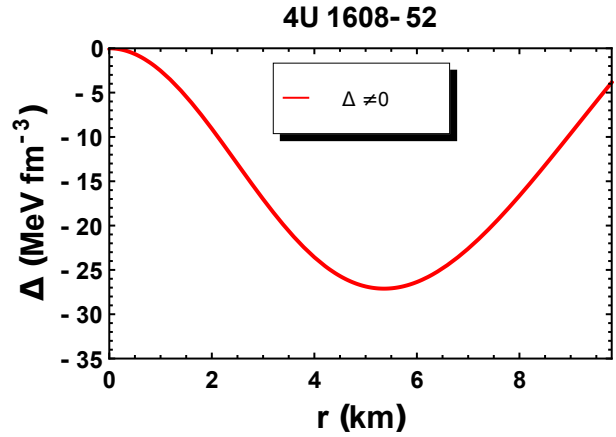}
\end{tabular}
\end{center}
\caption{Behavior of the anisotropy within the configuration with respect to the radial coordinate $r$. }\label{figani}
\end{figure}

The above choice for the anisotropy is physically reasonable, as at the center ($r=0$), anisotropy vanishes as expected. Fig.~\ref{figani} depicts the nature of anisotropy which clearly supports the regularity at the center. However the negative nature for the anisotropy leads us to some consequences as discussed subsequently. Similar profile of the anisotropic pressure can be observed in the work of Thirukkanesh et al.~\cite{TSD20}. As a limitation of our model, we can not generate the isotropic pressure condition from the specified anisotropic form. 

Also, this choice provides a solution to Eq.~(\ref{eq12}) in closed form. Substituting Eq.~(\ref{eq13}) in Eq.~(\ref{eq12}), we obtain
\begin{equation}
\frac{R\left( r(r^2 + R^2)^2 A_0'' - (r^2 + R^2)(2 r^2 +R^2) A_0'+ 2 r^3 A_0 \right) }{r(r^2 + R^2) A_0}=0.
\label{eq14}
\end{equation}

We obtain a simple solution to Eq.~(\ref{eq14}) as follows:
\begin{equation}\label{eq15}
A_0 = \left( C \sqrt{r^2 + R^2} + D \left( r^2 + R^2 \right) \right),
\end{equation}
where $C$ and $D$ are integration constants which will be obtained from the boundary conditions. 

With these choices of the metric potentials the matter density, radial pressure, transverse pressure and mass are obtained as
\begin{eqnarray}
8\pi\rho &=& \frac{r^2 + 3 R^2}{\left( r^2 + R^2 \right)^2 },\label{eq3b}\\
8\pi p_r &=& \frac{C \left(R^2 - r^2 \right) -D \left( r^2 - 3 R^2 \right) \sqrt{ r^2 + R^2 }}{\left( r^2 + R^2 \right)^2 \left( C + D \sqrt{ r^2 + R^2 } \right)},\label{eq4b}
\end{eqnarray}
\[8\pi p_t = \]
\begin{equation}
\frac{R^2 \left[ C \left(R^2 - r^2 \right) + D \sqrt{ r^2 + R^2 } \left( r^2 + 3 R^2 \right) \right] }{ \left( r^2 + R^2 \right)^3 \left(C + D \sqrt{ r^2 + R^2 } \right)},\label{eq5b}
\end{equation}
\begin{equation}
m(r) = \frac{r^3}{2 \left( r^2 + R^2 \right)}.\label{eq6b}
\end{equation}

Moreover, the gradients of the matter variables are obtained as
\begin{equation}
8\pi\frac{d\rho}{dr} = \frac{2 r}{\left(r^2+R^2\right)^2}-\frac{4 r \left(r^2+3 R^2\right)}{\left(r^2+R^2\right)^3},\label{eq7b}
\end{equation}

\[ 8 \pi \frac{dp_r}{dr} = \frac{2 r\text{C}^2 \left(r^2-3 R^2\right) \sqrt{r^2+R^2}}{\left(r^2+R^2\right)^{7/2} \left(\text{C}+\text{D} \sqrt{r^2+R^2}\right)^2}\]
\[+ \frac{2r \text{C} \text{D} \left(2 r^2-9 R^2\right) \left(r^2+R^2\right)}{\left(r^2+R^2\right)^{7/2} \left(\text{C}+\text{D} \sqrt{r^2+R^2}\right)^2}+ \]
\begin{equation} 
\frac{2r \text{D}^2 \left(r^2-7 R^2\right) \left(r^2+R^2\right)^{3/2}}{\left(r^2+R^2\right)^{7/2} \left(\text{C}+\text{D} \sqrt{r^2+R^2}\right)^2},\label{eq8b}
\end{equation}
\[ 8 \pi \frac{dp_t}{dr} = \]
\begin{equation}
\frac{R^2 \left( C \left(R^2-r^2\right)+ D \left(r^2+3 R^2\right) \sqrt{r^2+R^2}\right)}{\left(r^2+R^2\right)^3 \left(C+D \sqrt{r^2+R^2}\right)}.\label{eq9b}    
\end{equation}

\subsubsection{Case II: $\alpha = 1$}
The anisotropic factor now reduces to 
\begin{equation}
\Delta = \frac{r^2 (R^2 - r^2)}{(R^2 + r^2)^3}. \label{eqalpha1}
\end{equation}

Using the value of Eq.~(\ref{eqalpha1}), the master equation Eq.~(\ref{eq12}) reduces to the form
\[ \frac{\Big(r^2(r^2 + R^2)A''_0 - (2 r^2 + R^2)A'_0 \Big) R^2 (r^2 + R^2)}{r(r^2 + R^2)A_0}  + \]
\begin{equation}
\frac{2 r^4}{(r^2 + R^2)}= 0. \label{eqalpha2}
\end{equation}

Since the solution generated from Eq.~(\ref{eqalpha2}) are imaginary (See Appendix), we are excluding this discussions from the present paper. 

It is to note that in the original Finch-Skea paper~\cite{Finch} the exact solution in the presence of isotropy is available, i.e. for $p=p_r=p_t$ as such the isotropic pressure condition has been adopted. So, we are not repeating here the calculations based on the isotropic condition, however for the sake of comparison we have plotted graphs for both the cases which will help us to observe the effect of anisotropy in the present model.

\section{\label{Sec4} Exterior spacetime and boundary conditions}
The exterior space-time, spacetime outside the spherically symmetric configuration, for a non-radiating star is empty and is described by the exterior Schwarzschild solution as
\[ds^{2}=-\left(1-\frac{2M}{r}\right)dt^{2}+\left(1-\frac{2M}{r}\right)^{-1}dr^{2} + \]
\begin{equation}\label{extmS}
r^{2}\left(d\theta^{2}+\sin^{2}\theta d\phi^{2} \right),
\end{equation}
where $r>2M$, $M$ being the total mass of the stellar object. To study a compact stellar structure, the interior space-time metric (\ref{eq1}) must be matched smoothly to the exterior Schwarzschild spacetime metric Eq.~(\ref{extmS}) at the boundary of the star $r=b$. This condition is known as the continuity of the first fundamental form or Darmois-Israel condition~\cite{Darmois1927,Israel1966} and the continuity of the metric functions across the boundary $r=b$ yields
\begin{eqnarray}
A^2_0(b) &=& \left(1-\frac{2 M}{b}\right),\label{bc1}\\
B^2_0(b) &=& \left(1-\frac{2 M}{b}\right)^{-1}.\label{bc2}
\end{eqnarray}

The radial pressure drops to zero at a finite value of the radial parameter $r$, defined as the radius of the star. This is defined as the continuity of the second fundamental form and utilizing the condition $p_r(r=b)=0$, the radius of the star can be obtained as follows
\begin{equation}
 \left[-\frac{1}{b^2}+\frac{1}{B_0^2 b^2}+\frac{2A_0'}{b A_0B_0^2}\right]=0.\label{bc3}
\end{equation}

Fulfillment of continuity for both the first and the second fundamental forms is known as the junction condition and it is utilized to determine the constants for isotropic as well as anisotropic cases. Thus we have the constants in the forms:
\begin{eqnarray}
R &=& \frac{b \sqrt{b - 2 M}}{\sqrt{2 M}}, \label{R1}\\
C &=& \frac{M}{b^2} \left( 3 \sqrt{\frac{b-2M}{2M}}-\sqrt{\frac{2 M}{b-2M}}\right), \label{C1}\\
D &=& \sqrt{\frac{2 M^3}{b^7}}\left( \sqrt{\frac{2 M}{b-2M}}-\sqrt{\frac{b-2M}{2M}}\right), \label{D1}\\
G &=& \frac{ \sqrt{\frac{1}{\varphi}} \left( \sqrt{\varphi}~\cos \left(\sqrt{\varphi}\right) + \sin \left(\sqrt{\varphi}\right) \right)}{2 \sqrt{\varphi} \left( \cos \left(\varphi \right) + \sin \left(\varphi\right) \right)}, \label{G1}\\
H &=& \frac{ \sqrt{\frac{1}{\varphi}} \left( \sqrt{\varphi}~\sin \left(\sqrt{\varphi}\right) - \cos \left(\sqrt{\varphi}\right) \right)}{2 \sqrt{\varphi} \left( \cos \left(\varphi \right) + \sin \left(\varphi\right) \right)}. \label{H1}
\end{eqnarray}
where we take $\frac{b}{b - 2 M} = \varphi$.

\section{\label{Sec5} Physical analysis}
To study the physical features of the prescribed model we have considered the values from the pulsar $4$U~$1608-52$ as mass $=~1.57^{+0.30}_{-0.29}~M_\odot$ and radius $=~9.8 \pm 1.8$ km~\cite{roupas}. The values of the model parameters thus obtained are $R~=~10.3526$, $C~=~0.0535902$, $D~=~-0.000185649$, $G~=~0.328696$ and $H~=~0.305526$. We have analyzed the profile of the model analytically as well as graphically by considering the aforementioned dataset. However, the values of the model parameters for some other known compact objects are depicted in Table \ref{tab1}. \\

\begin{figure}[!htbp]
\begin{center}
\begin{tabular}{lr}
\includegraphics[width=8cm]{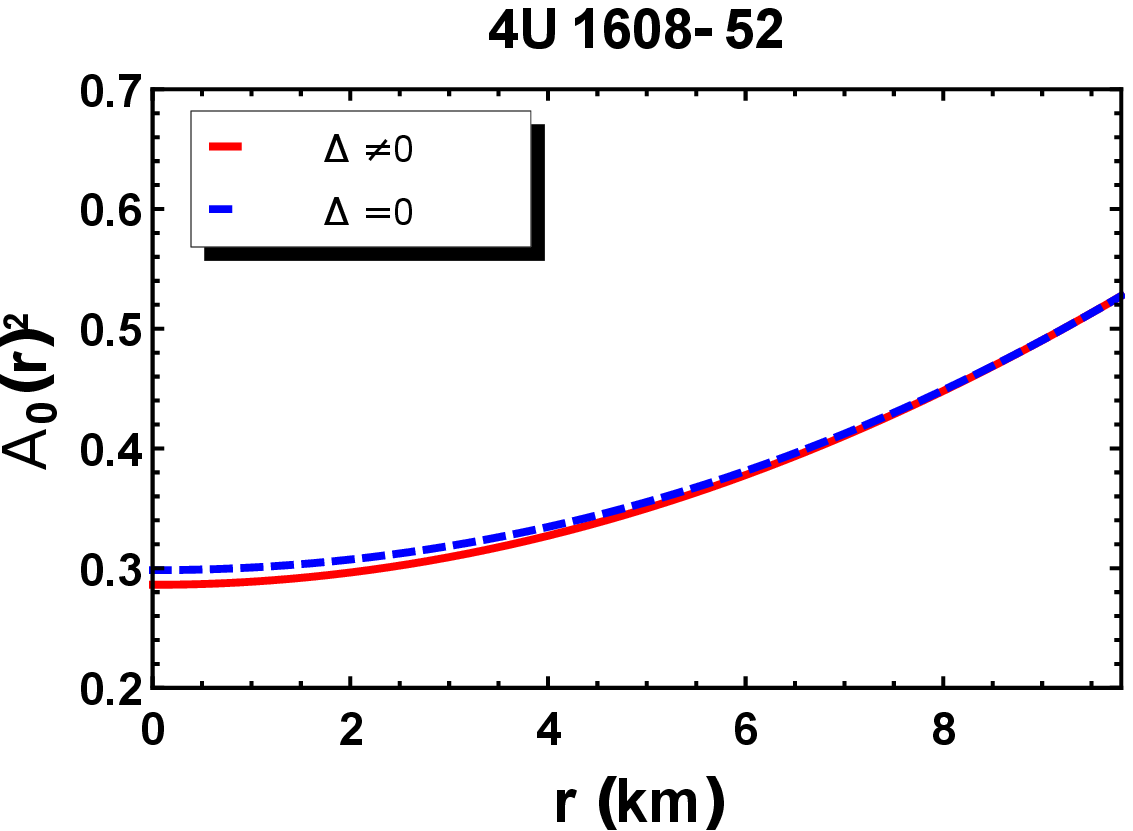}\\
\includegraphics[width=8cm]{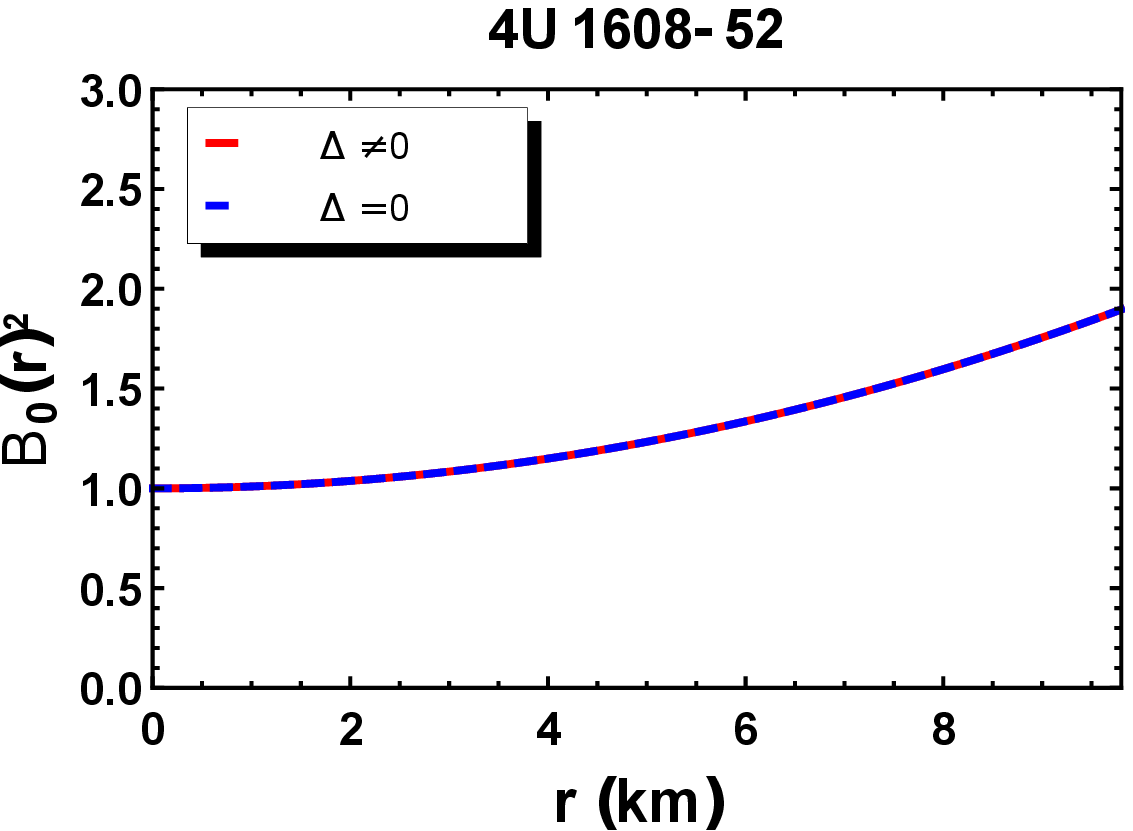}
\end{tabular}
\end{center}
\caption{Variation of the metric potentials $A^2_0(r)$ (above) and $B^2_0(r)$ (below) with the radial coordinate $r$. Here solid lines (red) represent the anisotropic case and the dashed lines (blue) represent the isotropic case. }\label{mp}
\end{figure}

\noindent 1. For any acceptable model, regularity of the solutions must be maintained. Analytically, the gravitational potentials should be free from any geometrical or physical singularity. Here,  $A^2_0(0)=R^2(C + DR)^2$=constant (for anisotropic case), $A^2_0(0)=0.5403(G-H)+0.84147(G+H)$= constant (for isotropic case) and $B^2_0(0)=1$, i.e., finite at the center ($r=0$) of the stellar configuration. Also one can easily check that $(A^2_0(r))'_{r=0}=(B^2_0(r))'_{r=0}=0$. These imply that the metric is regular at the center and well behaved throughout the stellar interior. Fig.~\ref{mp} exhibits the profile of the metric potentials within the stellar structure for both the anisotropic and isotropic cases.

\noindent 2. The central density, central radial pressure and central tangential pressure in this case are obtained as
\begin{equation}
8 \pi \rho(0) = {3 \over R^2}
\end{equation}

\begin{equation}
8 \pi p_r(0) = 8 \pi p_t(0) = \frac{C + 3 D R}{R^2 (C + D R)},
\end{equation}
for anisotropic case and
\begin{equation}
8 \pi p_r(0) = 8 \pi p_t(0) = \frac{(G+H) + 1.5574 (H -G)}{R^2 \left[(G -H) + 1.5574(G +H) \right]},
\end{equation}
for isotropic case.

Here $R$ being the curvature parameter, it is always positive, thus making the central density a positive quantity. For isotropic nature, both the pressures should be always equal whereas for anisotropic profile, the equality of the central values of both the radial pressure and tangential pressure depicts the absence of anisotropy at the center. The radial and tangential pressures at the center will be non-negative provided the chosen model parameters are all positive. Also according to Zeldovich's condition~\cite{Zeldovich1,Zeldovich2}, $p_r/\rho$ must be $\leq 1$ at the center. Therefore 
$$\frac{C + 3 D R}{3(C + D R )} \leq 1~{\text{and}}~\frac{(G + H) + 1.5574 (H -G)}{3\left[ (G-H) + 1.5574 (G+H) \right]}  \leq 1,$$ 
for anisotropic and isotropic cases respectively. 

\begin{figure}[!htbp]
\begin{center}
\begin{tabular}{lr}
\includegraphics[width=8cm]{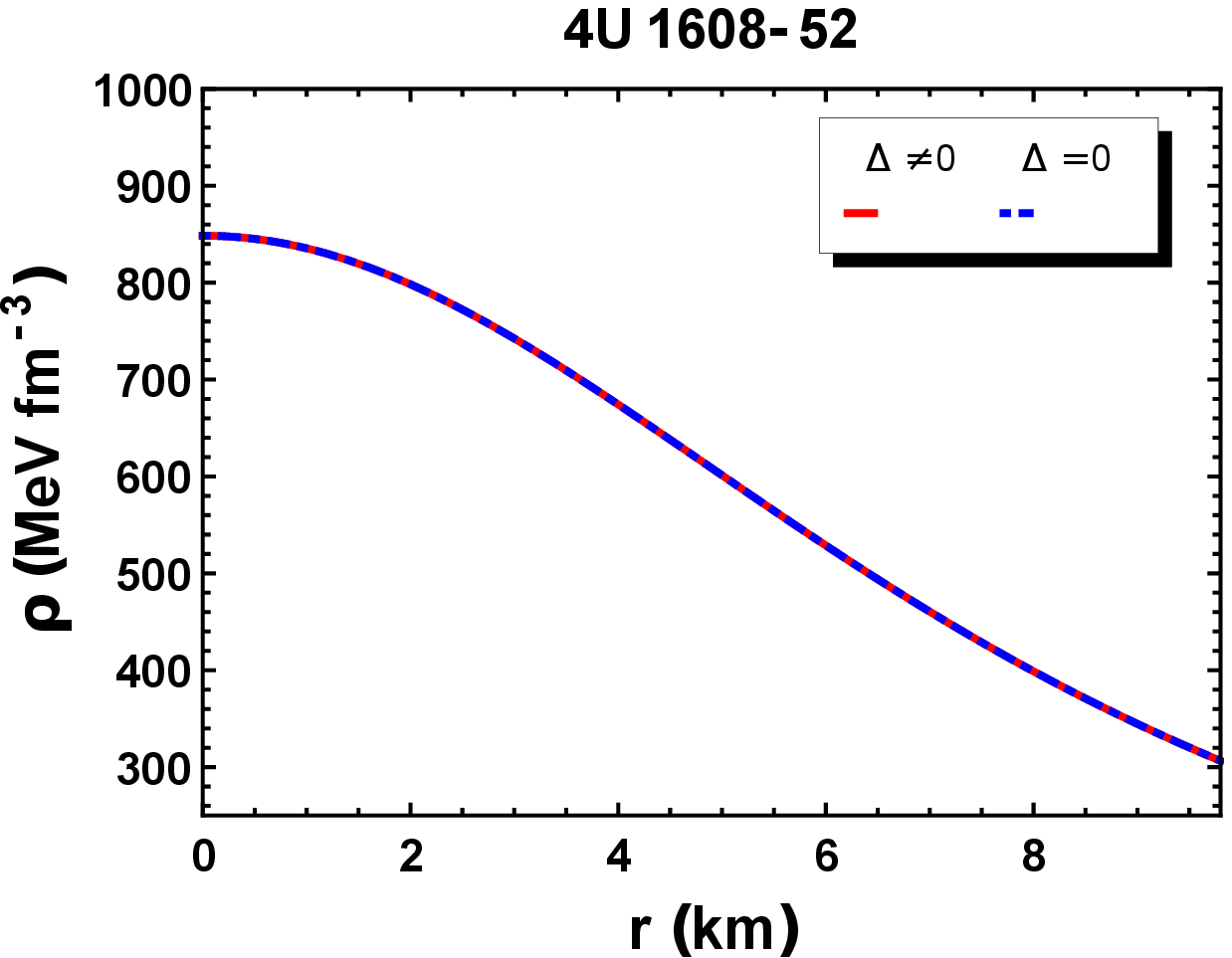}\\
\includegraphics[width=8cm]{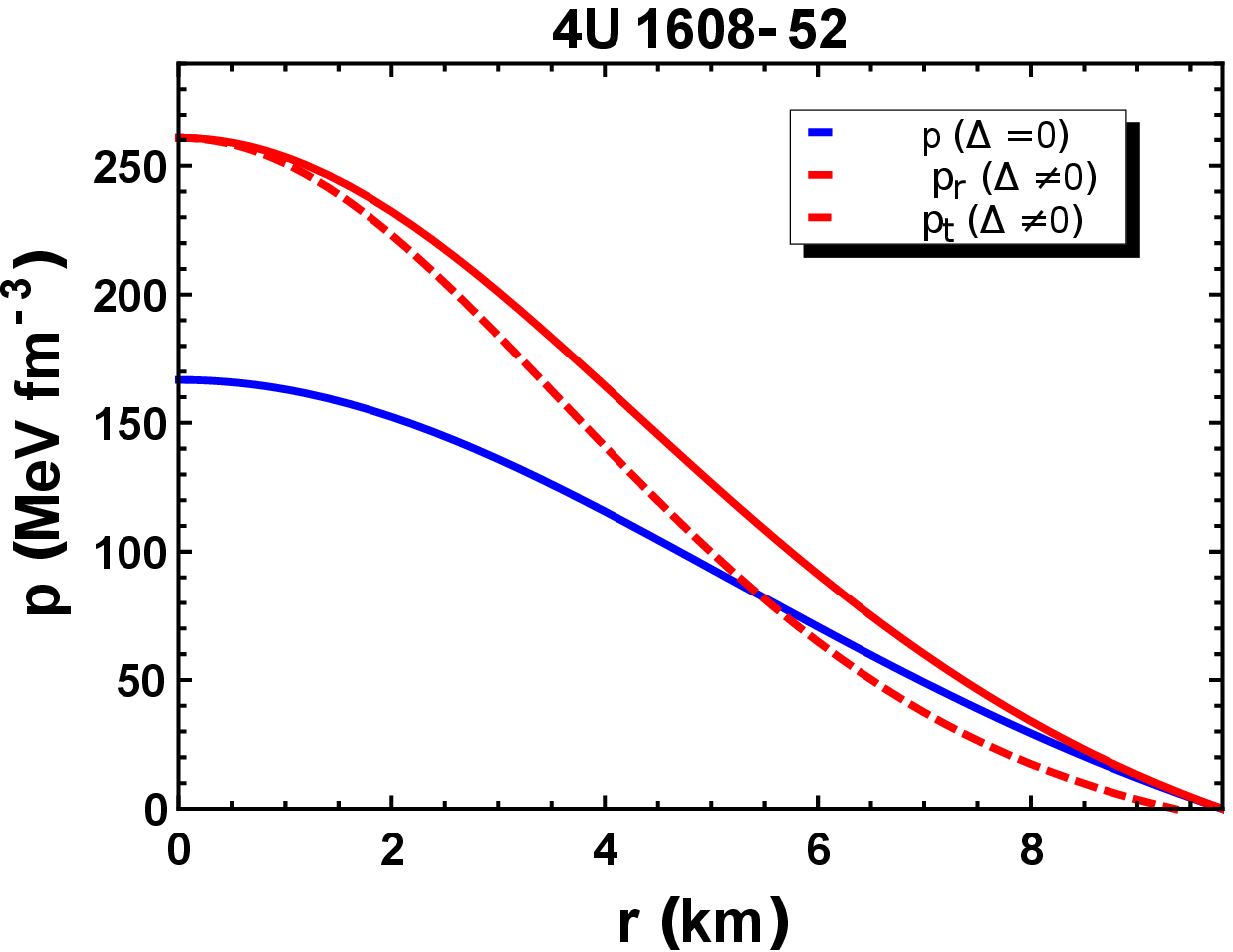}
\end{tabular}
\end{center}
\caption{Variation of the matter variables, density (above) and pressure (below) with the radial coordinate $r$. }\label{mvariable}
\end{figure}

The density $\rho$, radial pressure $p_r$ and transverse pressure $p_t$ are positive inside the structure and monotonically decreasing outward. Fig.~\ref{mvariable} supports the positive and monotonically decreasing behavior of the matter variables. However, since Fig.~\ref{mvariable} depicts that the transverse force is always lower than the radial one throughout the structure implying the attractive nature of the anisotropic force. This type of force is known to make the model less stable than the repulsive force.

\noindent 3. The density variation parameter $\lambda$ is defined as the ratio of the density at the surface to that at the center. Now for the prescribed model, the density variation parameter can be expressed as
\begin{equation}
\lambda = {\rho(b) \over \rho(0)}= \frac{R^2 (b^2 + 3 R^2)}{3 (b^2 + R^2)^2}. \nonumber
\end{equation}

Now for the fixed surface density of $2 \times 10^{14}~gm/cc$, Parui and Sarma~\cite{PS91} have deduced that minimum radius density (the ratio of the surface density to the radius of the star) is minimum for the $\lambda=0.68$. Based on this study, later Parui~\cite{Parui94} has generalized that for both the charged and uncharged neutron star of densities having $10^{15}$ and $10^{16}~gm/cc$, $\lambda_{max}$ becomes $0.68$ in each cases. For our model, considering the fixed surface density $2 \times 10^{14}~gm/cc$, the model parameter $R$ supports the value $0.673623$, the mass and radius become $0.6725~M_\odot$ and $0.346854$ km respectively for maximum limit of density variation parameter. However, the permissible value of $\lambda$ for several different stars of different surface densities are presented in Table~\ref{table2}. 

\noindent 4. The gradients of energy density, radial pressure and tangential pressures for anisotropic case are given in Eqs.~(\ref{eq7b})-(\ref{eq9b}).

\begin{figure}[!htbp]
\begin{center}
\begin{tabular}{lr}
\includegraphics[width=8cm]{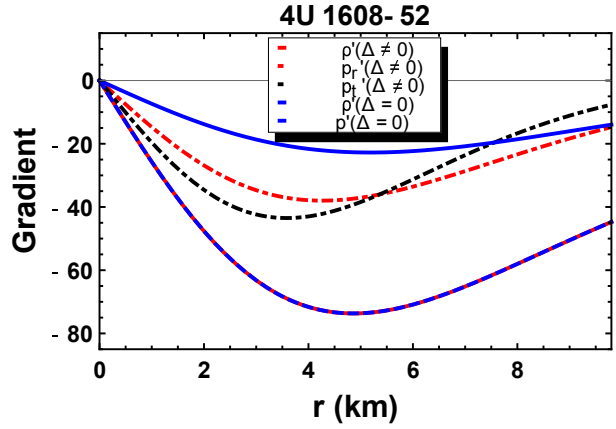}
\end{tabular}
\end{center}
\caption{Variation of gradients of the matter variables with the radial coordinate $r$.}\label{figgrad}
\end{figure}

The gradient of the density, radial pressure and tangential pressure are negative inside the stellar body are shown graphically in Fig.~\ref{figgrad}.

\noindent 5. The radial and transverse velocity of sound ($c=1$) are obtained as
\[v^2_{r} = \frac{-C^2(r^2 - 3 R^2)}{(r^2 + 5 R^2) (C + D \sqrt{r^2 + R^2})^2}  + \]
\[ \frac{D^2 (r^4 - 6 r^2 R^2 + 7 R^4)}{(r^2 + 5 R^2) (C + D \sqrt{r^2 + R^2})^2} + \]
\begin{equation}
\frac{ C D (- 2 r^4 + 7 r^2 R^2 + 9 R^4)}{\sqrt{r^2 + R^2} (r^2 + 5 R^2) (C + D \sqrt{r^2 + R^2})^2}, \label{velocity1}
\end{equation}
\[v^2_{t} = \frac{- 2 R^2 C^2 (r^2 - 2 R^2)}{(r^2 + R^2) (r^2 + 5 R^2) (C + D \sqrt{r^2 + R^2})^2} +\] \[\frac{2 R^2 D^2 (r^4 + 5 r^2 R^2 + 4 R^4)}{(r^2 + R^2) (r^2 + 5 R^2) (C + D \sqrt{r^2 + R^2})^2} +\]
\begin{equation}
 \frac{C D R^2 (- r^4 + 10 r^2 R^2 + 11 R^4)}{(r^2 + R^2)^{3 \over 2} (r^2 + 5 R^2) (C + D \sqrt{r^2 + R^2})^2},\label{velocity2}
\end{equation}
\[v^2 = \frac{(r^2 + R^2) \left(G + H \tan \Gamma \right) }{\Gamma (r^2 + 5 R^2)} \]
\[\times 
\frac{\left( - H (r^2 + R^2)+ G \Gamma (r^2 + 2 R^2) \right)}{\left[ R(G-H \Gamma) + (H + G \Gamma) \tan \Gamma \right]^2} + \]
\[ \frac{(r^2 + R^2) \left(G + H \tan \Gamma \right) }{\Gamma (r^2 + 5 R^2)} \]
\begin{equation}
\times  \frac{\left(G(r^2 + R^2)+ H \Gamma (r^2 + 2 R^2)\right) \tan \Gamma}{\left[ R(G-H \Gamma) + (H + G \Gamma) \tan \Gamma \right]^2},
\label{velocity3}
\end{equation}
where $\sqrt{1 + {r^2 \over R^2}} = \Gamma$. $v^2_{r}$ and $v^2_{t}$ are ${dp_r \over d\rho}$ and $dp_t \over d\rho$ respectively, described for the case of anisotropy. For isotropic case, $v^2$ denotes ${dp \over d\rho}$, $p$ being the pressure for prescribed model.

\begin{figure}[!htbp]
\begin{center}
\begin{tabular}{lr}
\includegraphics[width=8cm]{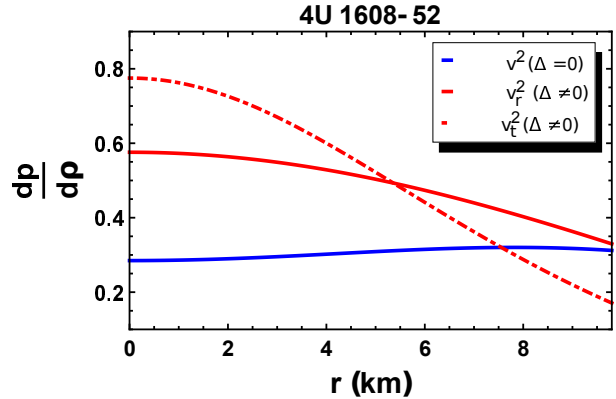}
\end{tabular}
\end{center}
\caption{Variation of the sound speeds with the radial coordinate $r$.}\label{figsound}
\end{figure}

In this model the speed of sound are smaller than $1$ in the interior of the star, i.e., $ 0 \leq \frac{dp_r}{d\rho} \leq 1 $, $ 0 \leq \frac{dp_t}{d\rho} \leq 1$ for anisotropic case and $0 \leq \frac{dp}{d\rho} \leq 1$ for isotropic case, which has been shown graphically in Fig.~\ref{figsound}.

\noindent 6. Energy Condition: The energy conditions play a crucial role to study the nature of the matter content in GR. The energy conditions are not physical constraints but are rather mathematically imposed boundary conditions on the matter variables. They restrict some contraction of the stress tensor at every spacetime point. The three main conditions studied here are: null energy condition (NEC), weak energy condition (WEC) and strong energy condition (SEC). The expressions for the energy conditions are described as follows:
\begin{eqnarray}
NEC_r  &:&  \rho(r) + p_r(r)\geq 0,~~ NEC_t  :  \rho(r) + p_t(r)\geq 0,\nonumber
\\
WEC_r  &:& \rho(r)\geq 0,~~ \rho(r) + p_r(r)\geq 0,\nonumber
\\
WEC_t  &:& \rho(r)\geq 0,~~ \rho(r) + p_t(r)\geq 0,\nonumber
\\
SEC  &:& \rho(r) + p_r(r) + 2p_t(r)\geq 0.  \nonumber 
\end{eqnarray}

\begin{figure}[!htbp]
\begin{center}
\begin{tabular}{lr}
\includegraphics[width=8cm]{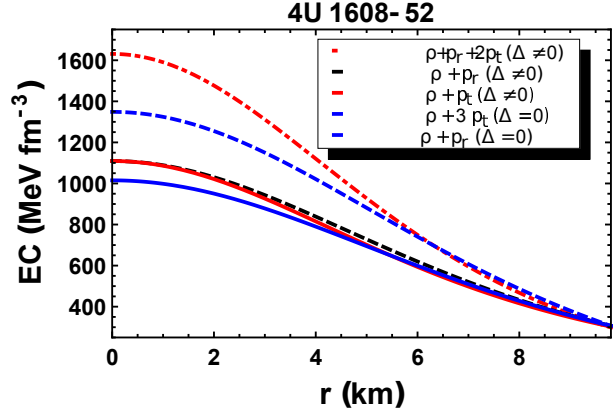}
\end{tabular}
\end{center}
\caption{Variation of various energy conditions with the radial coordinate $r$.}\label{figenergy}
\end{figure}

However, for an isotropic fluid sphere, the equality of the radial and the transverse pressure implies the expressions for the energy conditions as  $\rho + p_r  \geq 0$ and  $\rho + 3 p_t \geq 0$, throughout the stellar interior. These quantities are shown to remain positive throughout the compact sphere graphically in Fig.~\ref{figenergy}.  

\begin{figure}[!htbp]
\begin{center}
\begin{tabular}{lr}
\includegraphics[width=8cm]{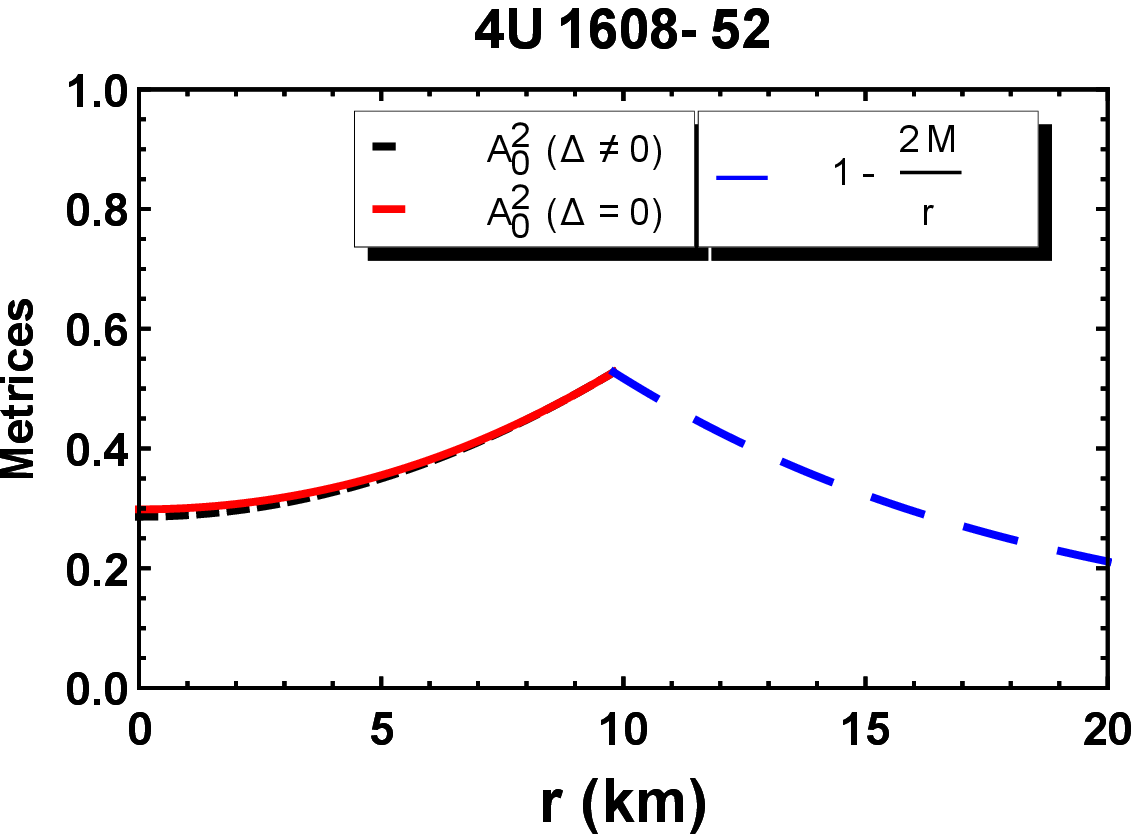}\\
\includegraphics[width=8cm]{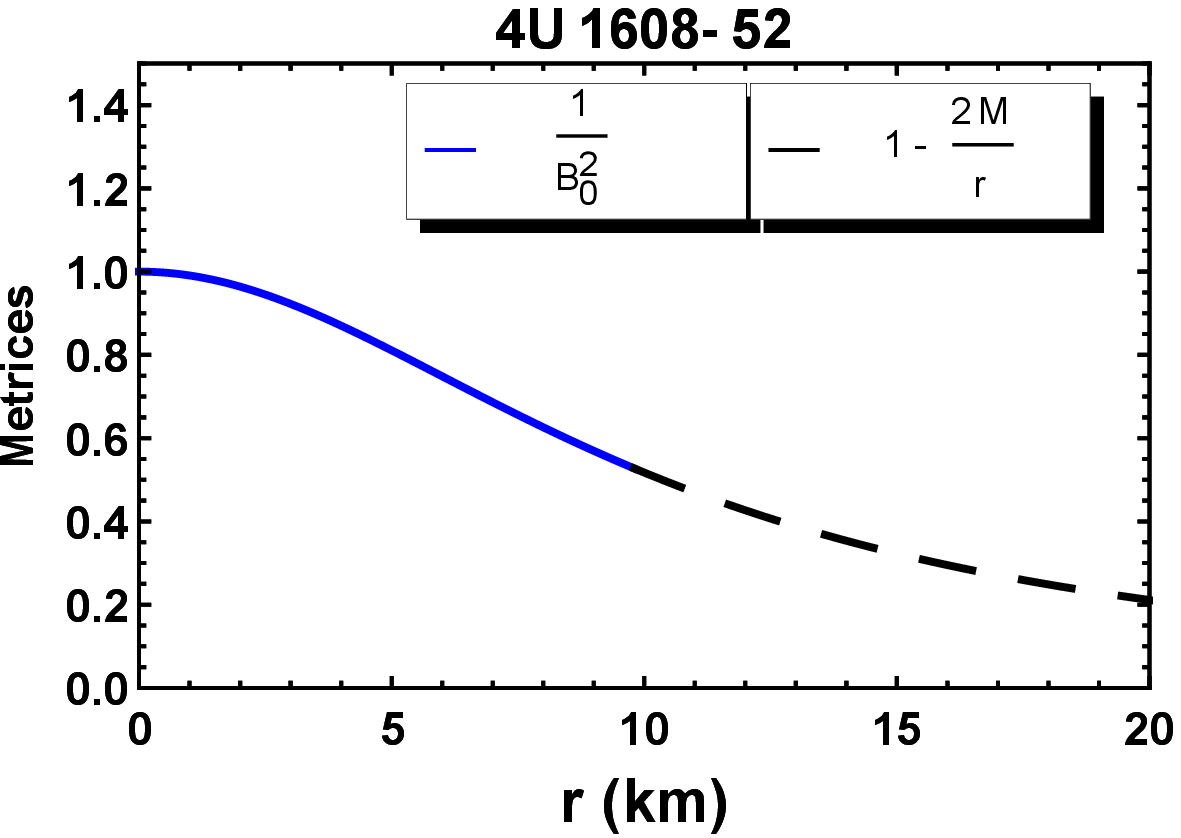}
\end{tabular}
\end{center}
\caption{Smooth matching of the metric potentials with Schwarzschild exterior solution at the boundary.}\label{figmatching}
\end{figure}

\noindent 7.  The smooth matching of the interior metric function with that of the Schwarzschild exterior at the boundary is shown graphically in Fig.~\ref{figmatching}. However, the formulation of the model constants obtained from the smooth matching at the boundary have been described in Sec.~\ref{Sec4}.  

\noindent 8. EOS parameter: The equation of state parameter is given by
\begin{equation}
\omega_r=\frac{p_r}{\rho};~
\omega_t=\frac{p_t}{\rho}.
\end{equation}

\begin{figure}[!htbp]
\begin{center}
\begin{tabular}{lr}
\includegraphics[width=8cm]{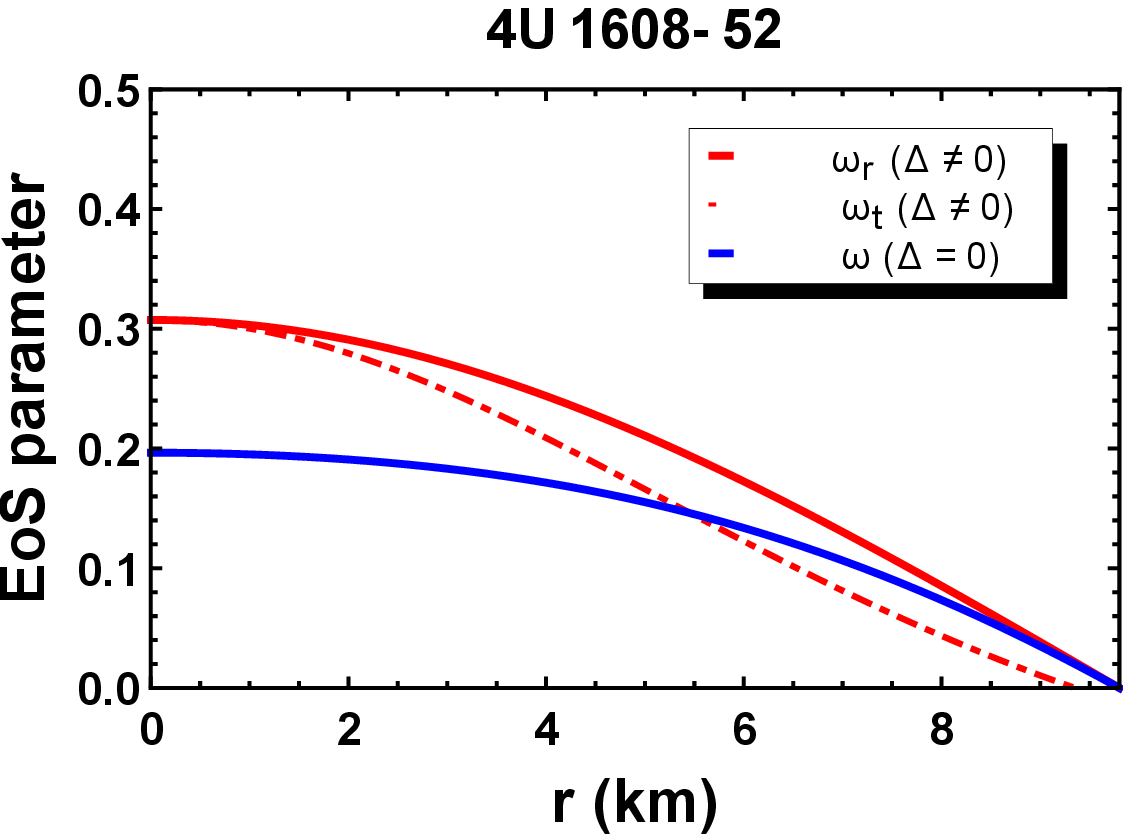}
\end{tabular}
\end{center}
\caption{Variation of EOS parameter inside the star with the radial distance for the anisotropic and isotropic pressures.}\label{figomega}
\end{figure}

To be non-exotic in nature the value of $\omega= p/\rho$ should lie within $0$ and $1$. The mathematical expressions for the EOS parameters can directly be obtained from the Eqs.~(\ref{eq3b})-(\ref{eq5b}). Graphically, our model is shown to satisfy the conditions $0\leq\omega_r\leq1$ and $0\leq\omega_t\leq1$ in Fig.~\ref{figomega}.

\begin{figure}[!htbp]
\begin{center}
\begin{tabular}{lr}
\includegraphics[width=8cm]{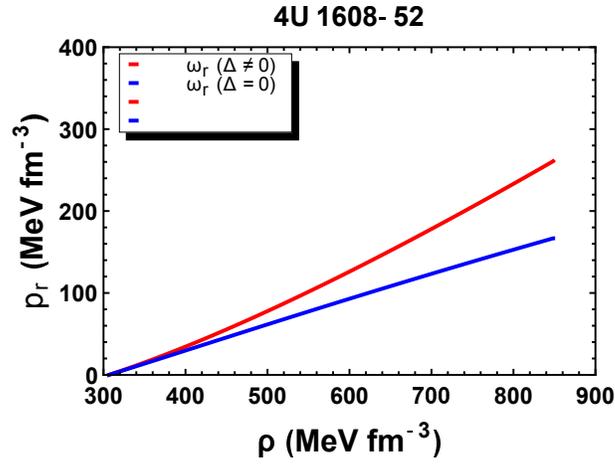}
\end{tabular}
\end{center}
\caption{Variation of the radial pressure with respect to density.}\label{figEOS}
\end{figure}

\begin{center}
\begin{table*}
\caption{Values of different model parameters corresponding to different known compact star}\label{tab1}
\begin{tabular}{|c|c|c|c|c|c|c|c|c|c|} \hline
Compact Star & Mass & Radius & R & C & D & G & H  \\
 &  ($M\odot$) &  (kms) &  & & & &   \\ \hline

SAX~J$1748.9$-$2021$~\cite{roupas} & $1.81^{+0.25}_{-0.37}$ & $11.7 \pm 1.7$ & $12.7697$ & $0.045989$ & $-0.000197$ & $0.344066$ & $0.302342$ \\ \hline

Cen~X-$3$~\cite{roupas} & $1.49 \pm 0.08$ & $9.17 \pm 0.13$ & $9.5572$ & $0.056641$ & $-0.000163$ & $0.322235$ & $0.306764$ \\ \hline

Vela~X-$1$~\cite{roupas} & $1.77 \pm 0.08$ &  $9.56 \pm 0.08$ & $8.7142$ & $0.06778$ & $0.000409$ & $0.255118$ & $0.316001$ \\ \hline

PSR~J$0030+0451$~\cite{miller} & $1.44^{+0.15}_{-0.16}$ &  $13.02^{+1.24}_{-1.06}$ & $18.7098$ & $0.045295$ & $-0.000407$ & $0.457842$ & $0.268874$  \\ \hline

\end{tabular}
\end{table*}
\end{center}

\begin{center}
  \begin{table*}
    \caption{Numerical values of the matter variables. Here $|_0$ and $|_b$ denote the values of the matter variables at the center and surface respectively. }
    \label{table2}
    \begin{tabular}{|c|c|c|c|c|}\hline
      \textbf{Compact Star} & \textbf{SAX~J$1748.9-2021$} & \textbf{Cen~X-$3$} & \textbf{Vela~X-$1$} & \textbf{PSR~J$0030+0451$} \\ 
      \textbf{Matter variables} & & & & \\ \hline
      
      \textbf{$\rho|_0$} & $557.659$ & $995.569$  & $1197.49$ & $259.772$ \\ \hline
      
      \textbf{$\rho|_b$} & $210.926$ & $352.714$ & $345.562$ & $136.949$ \\ \hline
      
      \textbf{$\lambda$} & $0.37823$ & $0.35428$ & $0.28857$ & $0.52718$ \\ \hline
      
      \textbf{$v^2_{r}|_0$} & $0.56381$ & $0.58101$ & $0.64117$ & $0.48499$ \\ \hline
      
       \textbf{$v^2_{r}|_b$} & $0.32863$ & $0.33014$ & $0.3392$ & $0.3286$ \\ \hline
       
        \textbf{$v^2_{t}|_0$} & $0.76306$ & $0.78027$ & $0.84043$ & $0.68426$ \\ \hline
        
        \textbf{$v^2_{t}|_b$} & $0.17529$ & $0.16937$ & $0.16679$ & $0.22002$ \\ \hline
        
        \textbf{$v^2|_0$} & $0.27526$ & $0.28898$ & $0.34177$ & $0.23487$ \\ \hline        
        
         \textbf{$v^2|_b$} & $0.30125$ & $0.31678$ & $0.37189$ & $0.23487$  \\ \hline
      
         \textbf{$(\rho + p_r + 2 p_t)|_0$} & $1048.86$ & $1931.26$ & $2559.74$ & $413.862$ \\ \hline
         
         \textbf{$(\rho + p_r + 2 p_t)|_b$} & $202.906$ & $345.893$ & $363.771$ & $123.771$ \\ \hline
         
         \textbf{$(\rho + 3 p)|_0$} & $865.771$ & $1598.53$ & $2156.35$ & $344.506$ \\ \hline
         
         \textbf{$(\rho + 3 p)|_b$} & $210.925$ & $352.714$ & $345.562$ & $136.949$ \\ \hline
         
         \textbf{$z|_b$} & $0.35627$ & $0.38586$ & $0.48443$ & $0.2183$ \\ \hline
      
    \end{tabular}
    \end{table*}
  \end{center}

\begin{center}
\begin{table*}
\caption{Comparison of the prescribed model with a neutron star model based on Walecka's relativistic mean field theory. Here densities are given in $10^{14}~gm/cc$.}\label{tab3}
\begin{tabular}{|c|c|c|c|c|c|c|c|} \hline
 \textbf{Mass ($M\odot$)} & \textbf{Radius (kms)} & \textbf{$\rho(0)$ (Walecka)} & \textbf{$\rho(0)$ (Model)} & \textbf{Error \%}  \\ \hline

$2.485$ & $11.271$ & $31.62$ & $23.74$ & $24.92$  \\ \hline

$2.543$ & $11.644$ & $25.12$ & $21.65$ & $13.81$  \\ \hline

$2.579$ & $12.027$ & $20.00$ & $19.29$ & $0.035$  \\ \hline

$2.583$ & $12.433$ & $15.85$ & $16.60$ & $-.047$  \\ \hline

$2.577$ & $12.521$ & $15.00$ & $15.98$ & $-0.065$  \\ \hline

$2.530$ & $12.798$ & $12.59$ & $13.847$ & $-0.0998$  \\ \hline

$2.387$ & $13.081$ & $10.00$ & $11.046$ & $-0.1046$  \\ \hline

$2.268$ & $13.167$ & $8.913$ & $9.65$ & $-0.08268$  \\ \hline

$2.119$ & $13.188$ & $7.943$ & $8.399$ & $-0.057$  \\ \hline

$1.919$ & $13.126$ & $7.080$ & $7.135$ & $-0.00776$  \\ \hline

$1.670$ & $12.949$ & $6.310$ & $5.936$ & $0.059$  \\ \hline

$1.400$ & $12.651$ & $5.623$ & $4.909$ & $0.126$  \\ \hline

$1.280$ & $12.486$ & $5.340$ & $4.5079$ & $0.1558$  \\ \hline

$1.123$ & $12.229$ & $5.012$ &  $4.0276$ & $0.196$  \\ \hline

$0.594$ & $11.033$ & $3.981$ & $2.514$ & $0.368$  \\ \hline

\end{tabular}
\end{table*}
\end{center}

\section{\label{Sec6} Stability analysis}

\subsection{Stability under different forces}
Modeling of a compact model requires to examine the stability of the model. The important characteristic to study the stability of any model is to check the equilibrium condition of the model by using TOV equation. This stability equation given by Tolman~\cite{tolman} and Oppenheimer and Volkoff~\cite{OV} symbolizes the internal structure of a spherically static symmetric compact object which is in equilibrium in the presence of anisotropy. The generalized TOV equation can be expressed as
\begin{equation}
-{M_G \over r}[\rho(r)+p_r(r)] {A_0(r) \over B_0(r)}-{dp_r(r) \over dr}+{2(p_t-p_r) \over r}=0,\label{force1}
\end{equation}
where $M_G(r)$ is the gravitational mass within the compact objects of radius $r$ which can be derived using Tolman-Whittaker mass formula and it is defined by
\begin{equation}
M_G(r) = \frac{r B_0(r) A_0'(r)}{A^2_0(r)}. \label{force2}
\end{equation}

Now, substituting the value of $M_G(r)$, Eq.~(\ref{force1}) can also be written as 
\begin{eqnarray}
-\frac{ A_0'(r) [\rho(r)+p_r(r)]}{A_0(r)}-{dp_r(r) \over dr}+{2(p_t-p_r) \over r}=0.\nonumber
\\\label{tov1}
\end{eqnarray}

Eq.~(\ref{tov1}) describe the equilibrium condition for the model under gravitational forces ($F_g$), hydrostatic forces ($F_h$) and anisotropic forces ($F_a$). The TOV equation can be expressed in a simple form as
\begin{equation}
F_g(r) + F_h(r) + F_a(r) =0,
\label{force3}
\end{equation}
where
\begin{eqnarray}
\text{Gravitational force}: F_g(r)& =& -\frac{ A_0'(r) [\rho(r)+p_r(r)]}{A_0(r)},\nonumber
\\
\text{Hydrostatic force}: F_h(r)& =& -{dp_r(r) \over dr},\nonumber
\\
\text{Anisotropic force}: F_a(r) &=& {2(p_t-p_r) \over r}.\label{force4}
\end{eqnarray}

\begin{figure}[!htbp]
\begin{center}
\begin{tabular}{lr}
\includegraphics[width=8cm]{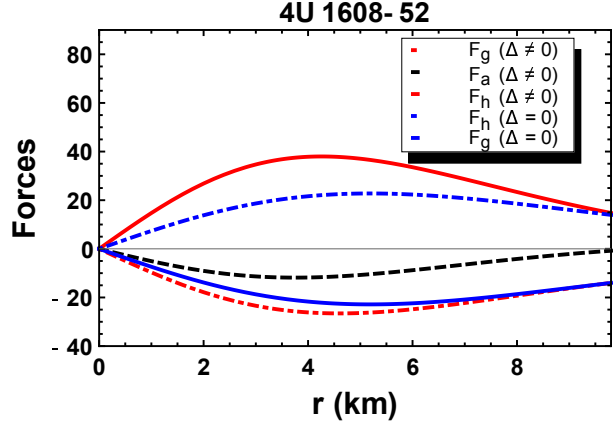}
\end{tabular}
\end{center}
\caption{Variations of different forces against $r$.}\label{figforce}
\end{figure}

The expressions mentioned in Eq.~(\ref{force4}) are examined graphically in the Fig.~\ref{figforce}. It portrays the stability of the model under various forces. It can clearly be seen that, to keep the model stable in the equilibrium, hydrostatic force should be much larger such that it can counterbalance the combined forces of gravitational and anisotropic forces. However, in the presence of isotropy, the model should be in the stable equilibrium if the negative gravitational force equalize the positive hydrostatic force.

\subsection{Stability under Causality Condition}
To examine the stability of a physically acceptable model, the velocity of the sound must be less than the light's velocity~\cite{LH,Abreu}. The sound velocity inside the compact star is expressed by
\begin{equation}
v_r(r)=\sqrt{{dp_r(r) \over d\rho(r)}},~~~v_t(r)=\sqrt{{dp_t(r) \over d\rho(r)}}.
\end{equation}

Since velocity of light $c = 1$, Thus  the causality condition becomes $0 \leq v_r(r), v_t(r) < 1$. Fig.~\ref{figsound} shows the fulfillment of the causality condition for both anisotropic and isotropic case.

\begin{figure}[!htbp]
\begin{center}
\begin{tabular}{lr}
\includegraphics[width=8cm]{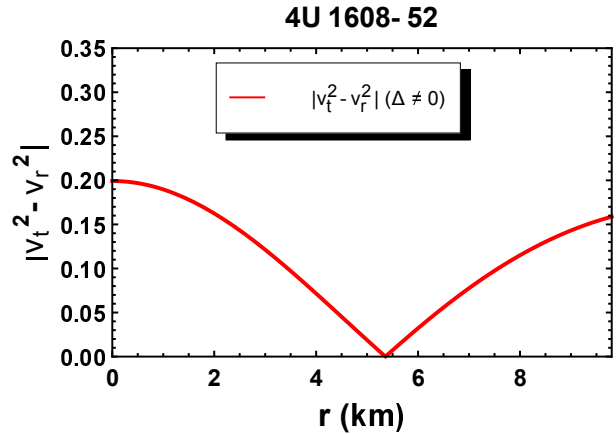}
\end{tabular}
\end{center}
\caption{The absolute difference of the sound speeds is plotted against $r$.}\label{figcausa}
\end{figure}

The stability of any compact object under the radial perturbation is investigated using Herrera's cracking concept~\cite{LH} and it is shown that to be a potentially stable model the absolute difference of sound speeds should be $\leq 1$~\cite{andreasson}. Fig. \ref{figcausa} portrays the stability condition for the prescribed model with anisotropic pressure. It is shown that the stability condition is satisfied by the model throughout the structure with anisotropic pressure.

\subsection{Stability under adiabatic index}
The adiabatic index, the ratio of the specific heats at the constant pressure and volume, is the quantity which incorporates all the basic characteristics of the equation of state on the instability formula and consequently consists the bridge between the relativistic structure of a spherical static object and the equation of state of the interior fluid~\cite{CCM}. Essentially it is a function of the baryon density and consequently exhibits the radial dependence on the instability criterion~\cite{Tooper}. Since the positive anisotropic factor may slow down the growth of instability which implies that the gravitational collapse occurs in the radial direction~\cite{MO}. Therefore, it is enough to study about adiabatic index only in the radial direction which is given mathematically as
\begin{eqnarray}
\Gamma_r(r) = {\rho(r)+p_r(r) \over p_r(r)}~{dp_r(r) \over d\rho(r)}.
\end{eqnarray}

\begin{figure}[!htbp]
\begin{center}
\begin{tabular}{lr}
\includegraphics[width=8cm]{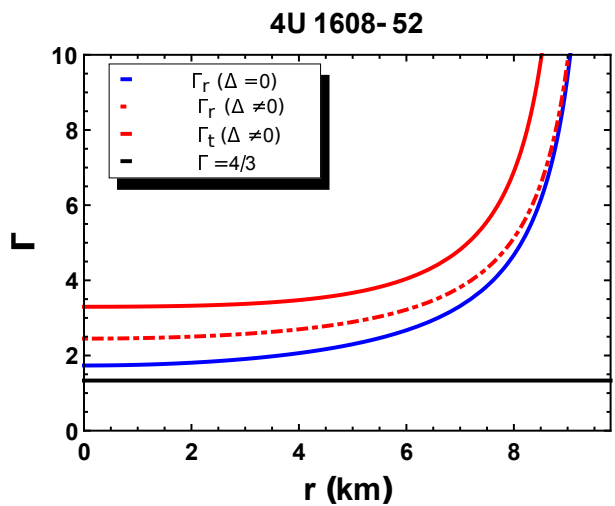}
\end{tabular}
\end{center}
\caption{The adiabatic indices plotted against $r$.}\label{figai}
\end{figure}

 We have checked the stability criteria graphically in Fig.~\ref{figai} and it can be seen that the model remains stable under both the anisotropic and isotropic pressures.
 
\subsection{Stability under the Harrison-Zeldovich-Novikov criterion}
One of the most important step to test the stability of the anisotropic compact star model is to check the stability of the mass of the model under Harrison~\cite{Harrison} and Zeldovich-Novikov~\cite{ZN} criterion. The general form is to test whether the mass is increasing with the increase of central density of a compact model. Mathematically, ${dM \over d\rho(0)}$ needs to be $<0$ to be stable structure, otherwise declared the model to be unstable. For our model the mass can be written in the form central density as following:
\begin{eqnarray}
M(\rho(0)) &=& \frac{b^3 \rho(0)}{2 (b^2 \rho(0) + 3)}, \label{cenmass}\\
\frac{dM}{d\rho(0)} &=& \frac{3 b^3}{2 (b^2 \rho(0) +3)^2}. \label{dcenmass}
\end{eqnarray}

\begin{figure}[!htbp]
\begin{center}
\begin{tabular}{lr}
\includegraphics[width=8cm]{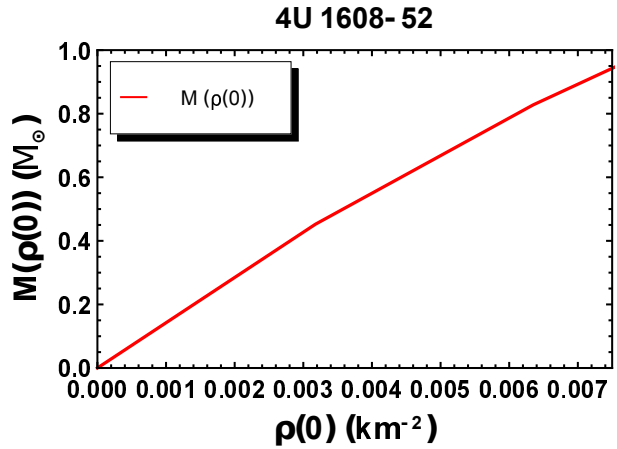}
\end{tabular}
\end{center}
\caption{Variation of the mass with respect to the central density.}\label{fighzn}
\end{figure}

\begin{figure}[!htbp]
\begin{center}
\begin{tabular}{lr}
\includegraphics[width=8cm]{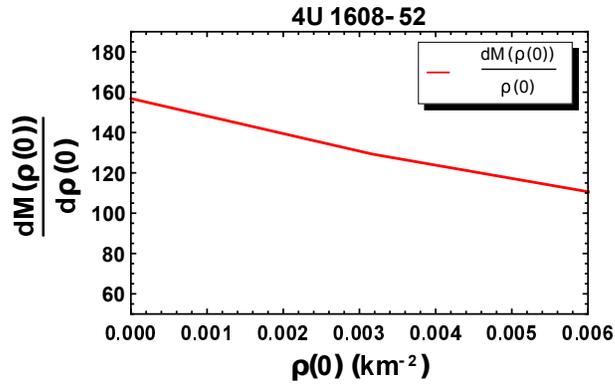}
\end{tabular}
\end{center}
\caption{Variation of the gradient of the mass against the central density.}\label{figdhzn}
\end{figure}

The profile of mass and the gradient of mass in the form of central density has been depicted in Figs.~\ref{fighzn} and \ref{figdhzn} respectively. It can clearly be seen that ${dM \over d\rho}$ is positive throughout the stellar configuration making it stable.

\section{\label{Sec7} Mass-Radius relationship and redshift}

\subsection{Mass function and mass-radius relationship}
\begin{figure}[!htbp]
\begin{center}
\begin{tabular}{lr}
\includegraphics[width=8cm]{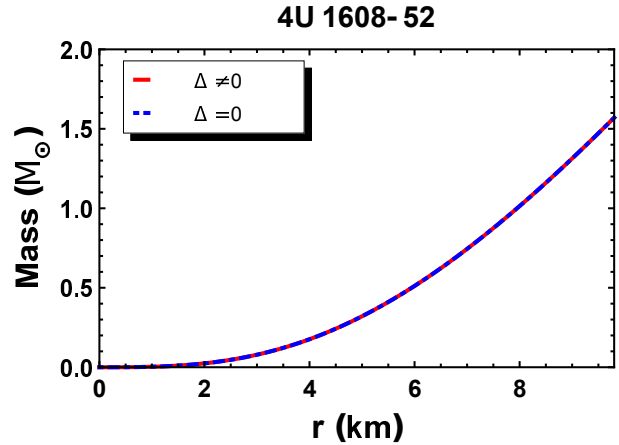}
\end{tabular}
\end{center}
\caption{The profile of the mass function plotted against $r$.}\label{fimassfn}
\end{figure}

The mass function for the model is given in Eqs~(\ref{eq6b}). Since $\lim_{b \to 0} m(b) = 0$, so the mass function is regular at the center of the structure. Also Fig.~\ref{fimassfn} depicts the positive and monotonically increasing nature of the mass function with respect to the radial coordinate. 

\begin{figure}[!htbp]
\begin{center}
\begin{tabular}{lr}
\includegraphics[width=8cm]{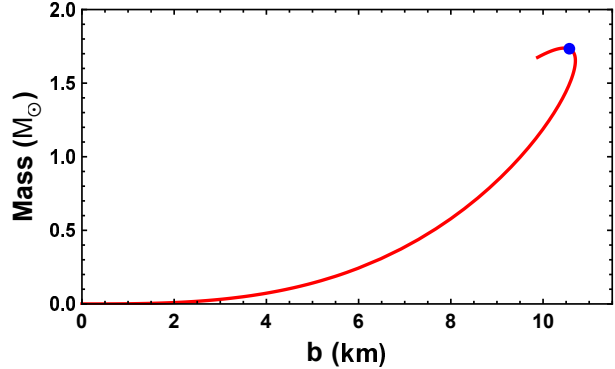}
\end{tabular}
\end{center}
\caption{The mass-radius relationship for the prescribed compact model. The solid circle represents the maximum mass attained by the model.}\label{figmr}
\end{figure}

The mass-radius relationship for the model is plotted in Fig.~\ref{figmr} and the maximum mass obtained is $1.731~M_\odot$ corresponding to the radius $10.56~km$ considering the fixed surface density $5.5 \times 10^{14}~gm/cc$. Since we know from the works of Sharma et al.~\cite{SDT} and Sunzu et al.~\cite{SMR}, the mass radius relationship is not affected by the pressure anisotropy. Hence the obtained mass-radius relation can be obtained both for anisotropy and isotropy cases. Also for spherically symmetric stable structure, Buchdahl limit~\cite{Buchdahl} needs to be satisfied, i.e. ${2M \over b}$ must be less than ${8 \over 9}$. 

\subsection{Equation of State}
To study cold high-density matter, compact objects act as the natural laboratories and such behavior is governed by the relation between pressure and density, known as the Equation of State (EOS). We can study the mass and radius as well as other macroscopic properties such as moment of inertia and tidal deformability of a compact star. The variation of the pressure with the density is plotted in Fig.~\ref{figEOS}. It can be seen that the anisotropic pressure generates more stiff EOS than that of isotropic pressure. Same results can be concluded fron Fig.~\ref{figsound} also. Since we know the stiffness of EOS can observed from the variation of the sound speed in stellar medium. 

Now stiffer EOS lead to larger tidal deformability with the anistropic pressure and the presence of anisotropy can reduce the value of the dimensionless tidal deformability by a significant amount for a given mass~\cite{BB}. However, Biswas and Bose~\cite{BB} have exclusively studied the case for positive anisotropy. 

\subsection{Mass-central density relationship}
\begin{figure}[!htbp]
\begin{center}
\begin{tabular}{lr}
\includegraphics[width=8cm]{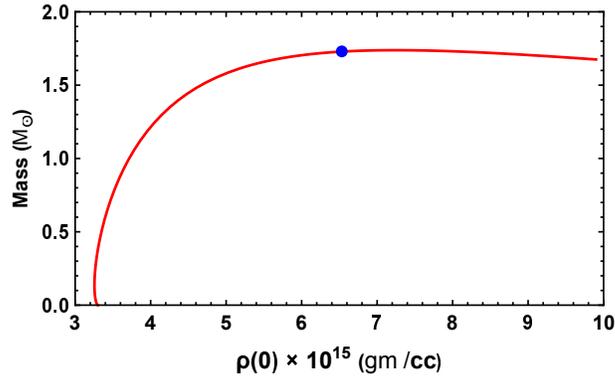}
\end{tabular}
\end{center}
\caption{The mass-central density relationship for the prescribed compact model. Here solid circle denotes the maximum mass for the model.}\label{figmrho}
\end{figure}

The stability of any model depends on the variation of mass with the central density and it is known as Harrison-Zeldovich-Novikov criterion (which is discussed in the previous sub-section). This criterion states that the model is stable in stellar system only if the mass of the model is increasing with the increase of central density. In Fig.~\ref{figmrho} the increasing nature of mass with respect to central density is quite evident. Also it is to be noted that the central density does not vanish for the absence of mass. For our prescribed model, the maximum mass of the model corresponds to the central density $6.537 \times 10^{15}~gm/cc$.

\subsection{Radius-central density relationship}
To examine any viable model it is important to investigate the central density against the radius along with the mass of the model. 

\begin{figure}[!htbp]
\begin{center}
\begin{tabular}{lr}
\includegraphics[width=8cm]{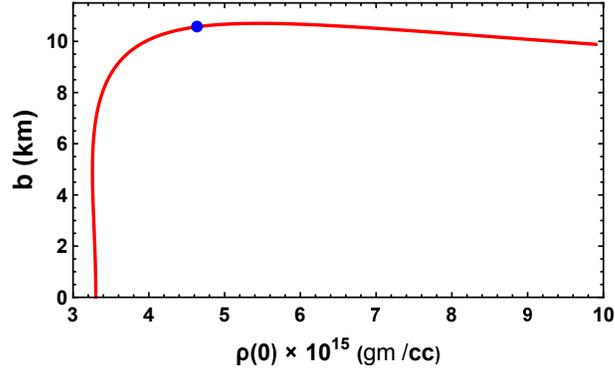}
\end{tabular}
\end{center}
\caption{The radius-central density relationship for the prescribed compact model. The solid circle represents the radius for which the maximum mass is attained.}\label{figrcenden}
\end{figure}

The radius-central density relationship is plotted in Fig.~\ref{figrcenden}. It can be observed that central density increases with the increase of radius of the model. Here the maximum central density corresponding to the radius $10.56~km$ is obtained as $4.63 \times 10^{15}~gm/cc$.

\subsection{Surface redshift}
The compactness of the model is defined by a dimensionless parameter $u(r)= {m(r) \over r}$. According to Buchdahl limit~\cite{Buchdahl}, the compactness of a model should be less than $0.444$ to be a stable structure. For our model we have the compactness of our model as $0.2417$ indicating the fulfillment of the Buchdahl condition.

The surface redshift of a spherically symmetric compact object is defined by
\begin{equation}
z = \frac{1}{\sqrt{1 - 2 u(r)}} - 1, \label{eqz}
\end{equation}
where $u(r) = m(r) /r$ is the compactness parameter for the model.

\begin{figure}[!htbp]
\begin{center}
\begin{tabular}{lr}
\includegraphics[width=8cm]{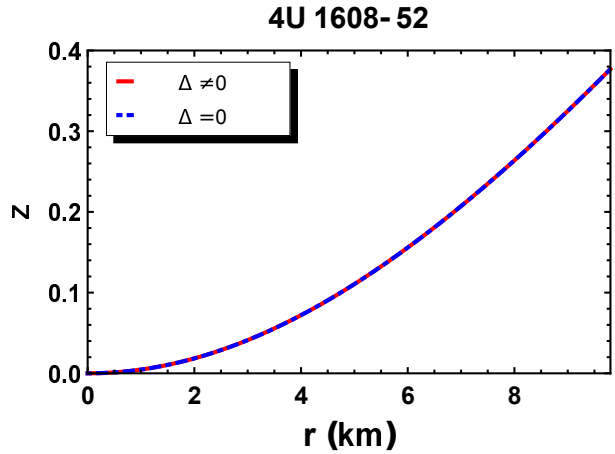}
\end{tabular}
\end{center}
\caption{The surface redshift plotted against $r$.} \label{figred}
\end{figure}

The surface redshift is plotted against the radial coordinate in Fig.~\ref{figred}.

\section{\label{Sec8} Discussions and conclusion}
We have analyzed the field equations adopting the Finch-Skea ansatz~\cite{Finch} and have obtained a solution that describes a compact model with negative anisotropic pressure. Some salient features of the solution are described as follows: 
\begin{itemize}
\item[(1)] The profile for anisotropic pressure of the model have been studied in Fig.~\ref{figani}. Though the anisotropic parameter satisfies the regularity at the center ($r~=~0$), it is shown to portray negative nature throughout the stellar structure, making the anisotropic force acting on the model to be attractive and which is proven to make the model less supportive against the gravitational collapse.

\item[(2)] All the matter variables for the compact model satisfy the physical requirements to be a stable model. Energy density and the profile of pressures in the presence of anisotropy as well as isotropy are plotted in Fig.~\ref{mvariable}. In presence of anisotropy, transverse pressure is observed to be less than the radial one throughout the structure.

\item[(3)] Smooth matching of the interior solutions with the Schwarzschild exterior solutions at the boundary in Fig.~\ref{figmatching} helps to generate the general form for the constants which further provides an outline of the compact stellar model.

\item[(4)] Variation of gradients of matter variables are shown to be negative throughout the star with zero gradients at the center.

\item[(5)] The causality condition is satisfied by the variation of the sound speed as shown in Fig.~\ref{figsound}. Also the absolute difference of the sound speeds are plotted in Fig.~\ref{figcausa}, implying that the model does not satisfy the stability condition by Herrera Cracking concept.

\item[(6)] Fulfillment of various energy conditions by the model in presence of anisotropy and isotropy are shown in Fig.~\ref{figenergy}.

\item[(7)] The stability of the model under the effect of TOV equation is shown in Fig.~\ref{figforce}. The model is shown to remain in static equilibrium if the hydrostatic force neutralize the combined effect of the anisotropic and gravitational forces. 

\item[(8)] The monotonically increasing nature of the mass function and the surface redshifts are plotted in Figs.~\ref{fimassfn} and \ref{figred} respectively, which support the physical viability of the prescribed model.

\item[(9)] The maximum mass, obtained for the prescribed model, is $1.731~M_\odot$ corresponding to the radius $10.56$ $km$, which is stable as per Buchdahl limit. Also the radius-central density relationship depicted in Fig.~\ref{figrcenden} illustrate that the central density increases with the increase of radius of the model. The central density is obtained to be $4.63 \times 10^{15}$ $gm/cc$ corresponding to the radius such that maximum mass is obtained in Fig.~\ref{figmr}.
\end{itemize}

We have also represented tables for a comparative study considering some well known stars. Table \ref{tab1} depicts the value of the model parameter while Table \ref{table2} exhibits the values of the matter variables for both anisotropic and isotropic scenario. Obviously the obtained solution is reduced to the solution obtained by Finch-Skea~\cite{Finch} by assuming zero anisotropy. 

However, as a final comment we would like to point out that if some physical constraint such as an equation of state, conformal geometry, embedding, etc were invoked then the study would take on a more meaningful flavor. This aspects may be considered seriously in a future project.

\subsection*{Appendix}
The solutions for Eq.~(\ref{eqalpha2}) are obtained using technical computing system as
\begin{equation}
A_0(r) = 2^{-1 \over 4} (-R)^{2n+3 \over 2} \mathit{s}^{3 \over 2} \left[ M \mathcal{I}_n(\mathit{s}) - N (-1)^n \mathcal{K}_n(\mathit{s}) \right], \nonumber
\end{equation}
where $n = {\sqrt{17} \over 2}$, $\mathit{s} = \sqrt{\frac{-2(r^2 + R^2)}{R^2}}$ and $\mathcal{I}$, $\mathcal{K}$ are the modified Bessel's functions of first and second order respectively.

If we try to solve the equations in other approach namely, by transformation we obtain the following results.

If we use Durgapal-Banerjii transformation~\cite{Durgapal1983}, i.e., set $x = {r^2 \over R^2}$, $Z(x) = {1 \over B^2_0(r)}$ and $A^2 y^2 (x)= A^2_0(r)$, we get the the field equations along with the anisotropic factor to be transformed as
\begin{eqnarray}
8 \pi \rho &=& \frac{1 - Z(x)}{x R^2}- {2 Z' \over R^2}, \nonumber \\
8 \pi p_r &=& \frac{Z-1}{x R^2} + \frac{4 Z y'}{R^2 y}, \nonumber \\
\Delta (x) &=& \frac{x(1 - x)}{R^2(1 +x )^3}, \nonumber \\
8 \pi p_t &=& 8 \pi p_r + \Delta (x), \nonumber
\end{eqnarray} 
where (') denotes differentiation of the respective function with respect to $x$. Now combining all the above equations and using the transformation $Z(x) = \frac{1}{1 +x}$, we get a second order ODE as
\begin{equation}
4(1 + x)^2 y'' -2 (1 + x)y' + (1 +2 x)y =0. \label{eqx}
\end{equation}
Now we try to solve Eq.~(\ref{eqx}) using some known methods.\\

If we again use transformations as $1+x = V$ and $y = Y V^{3 \over 4}$ on Eq.~(\ref{eqx}), we get the new transformed ODE as
\begin{equation}
V^2 {d^2Y \over dV^2} + V {dY \over dV} + \left({V \over 2} - 1\right)Y=0, \nonumber
\end{equation}
which cannot be simplified in Bessel's form of differential equations. Now as per series solutions, we cannot further use Frobenius Method to solve the above ODE as $\left( {1 \over V}\right)$ is not analytic at $V=0$.

\section*{Acknowledgement}
SD, KC and SR are thankful to the authority of Inter-University
Centre for Astronomy and Astrophysics, Pune, India for providing
them Visiting Associateship under which a part of this work was
carried out.

\end{document}